\UseRawInputEncoding
\documentclass[pre,aps,preprint,amsmath,superscriptaddress,showkeys]{revtex4}

\usepackage{graphicx}
\usepackage{dcolumn}
\usepackage{bm}

\begin{document}

\title{Spatiotemporal Multiphysics Metamaterials with Continuously Adjustable Functions}

\author{Min Lei}
\affiliation{Department of Physics, State Key Laboratory of Surface Physics, and Key Laboratory of Micro and Nano Photonic Structures (MOE), Fudan University, Shanghai 200438, China}

\author{Liujun Xu}\email{ljxu@gscaep.ac.cn}
\affiliation{Graduate School of China Academy of Engineering Physics, Beijing 100193, China}

\author{Jiping Huang}\email{jphuang@fudan.edu.cn}
\affiliation{Department of Physics, State Key Laboratory of Surface Physics, and Key Laboratory of Micro and Nano Photonic Structures (MOE), Fudan University, Shanghai 200438, China}

\date{\today}

\begin{abstract}
	Emerging multiphysics metamaterials offer a distinct possibility for regulating complex physical processes. However, two severe constraints still lower their functionality and tunability. First, multiphysics functionality is fixed once structures and materials are prepared, i.e., one functionality for one physical field. Second, continuous tunability is unavailable in multiphysics fields because parameters are hard to change on demand. Here, we propose the concept of spatiotemporal multiphysics metamaterials by delicately considering the temporal dimension. The spatiotemporal feature leads to multiple functions for each physical field and their continuous switching. We achieve flexible thermal and electric function switching between cloaking, sensing, and concentrating based on rotatable checkerboard structures with different rotation times, material composition, and geometric shapes. Real-time thermal and electric functions are theoretically predicted and confirmed by simulations. These results provide a promising spatiotemporal platform for realizing adaptive and intelligent multiphysics field manipulation.
\end{abstract}


\maketitle

\section{Introduction}
Physical field control in complex scenarios is significant for various areas, such as power generation~\cite{EES21}, radiative cooling~\cite{NP22}, biomedical engineering~\cite{MTA22}, and energy management~\cite{MTP21-2}. Metamaterials~\cite{AM21,MTP22} provide a fabulous paradigm to manipulate diverse physical fields like optics~\cite{APR22}, acoustics~\cite{NRM22}, thermotics~\cite{PR21,NRP23}, and fluidics~\cite{PRL19,PNAS22}. Exciting functions include cloaking, sensing, and concentrating, which make the physical field intensity in the working region far smaller, equal, and larger than in the background~\cite{PR21,MTP21-1,MTP22-1,MTP22-2}. With higher application demands, metamaterial design goes beyond the restriction of ``one metamaterial for controlling one physical field,'' and multiphysics metamaterials demonstrate significant potential in realizing ``one metamaterial for regulating multiple physical fields.'' This trend leads to the simultaneous control of electromagnetic plus acoustic fields ~\cite{ACS19,AOM2020}, conductive plus convective or radiative fields~\cite{NM19,PRAP20,LiAM20,IJHMT21}, and thermal plus electric fields~\cite{JAP2010,PRL2014-3,PRX2014,AM15,OE2015,SR2017}. However, existing multiphysics metamaterials still have severe limitations. On the one hand, these conventional schemes have only fixed functions due to the restriction of static structures and materials. On the other hand, continuous controllability is unavailable because multiphysics parameters are hard to change simultaneously and at will. Therefore, it is still a big challenge for multiphysics metamaterials to achieve multiple functions for each physical field, let alone their continuous switching.

Metamaterial design containing the temporal dimension~\cite{eLight22,AP22} has recently drawn intensive attraction for adding a promising degree of freedom. Regarding fundamental physics, the spatiotemporal modulation contributes to unexpected nonreciprocal phenomena~\cite{AM2019,AP2022-1,SA21,AM22,PRL2022-1,PRL2022-2,IJHMT2022} and intriguing topological transport~\cite{AOM2021,NC22,NP2022}. Besides, the spatiotemporal approach yields practical applications in adaptive thermal camouflage~\cite{AM2017} and real-time digital coding~\cite{NC2018,NC2019}. However, these spatiotemporal schemes apply only to a single physical field. A tricky problem for extending them to multiphysics fields is that parameters of different physical fields are highly inconvenient to regulate on demand due to structure and material restrictions.

Here, we propose the concept of spatiotemporal multiphysics metamaterials to control thermal and electric fields simultaneously. The temporal dimension is delicately introduced by a rotatable checkerboard structure whose geometric configuration can be continuously tuned over time. Thus, we can achieve the flexible control of thermal and electric conductivities and realize various functions in thermal and electric fields. Our spatiotemporal multiphysics metamaterials feature flexible switching between three (or five) function combinations via two (or four) constituent materials; see Fig.~1a. These results provide an unprecedented possibility for flexible and intelligent multiphysics field control.

\begin{figure}
	\centering
	\includegraphics[width=\linewidth]{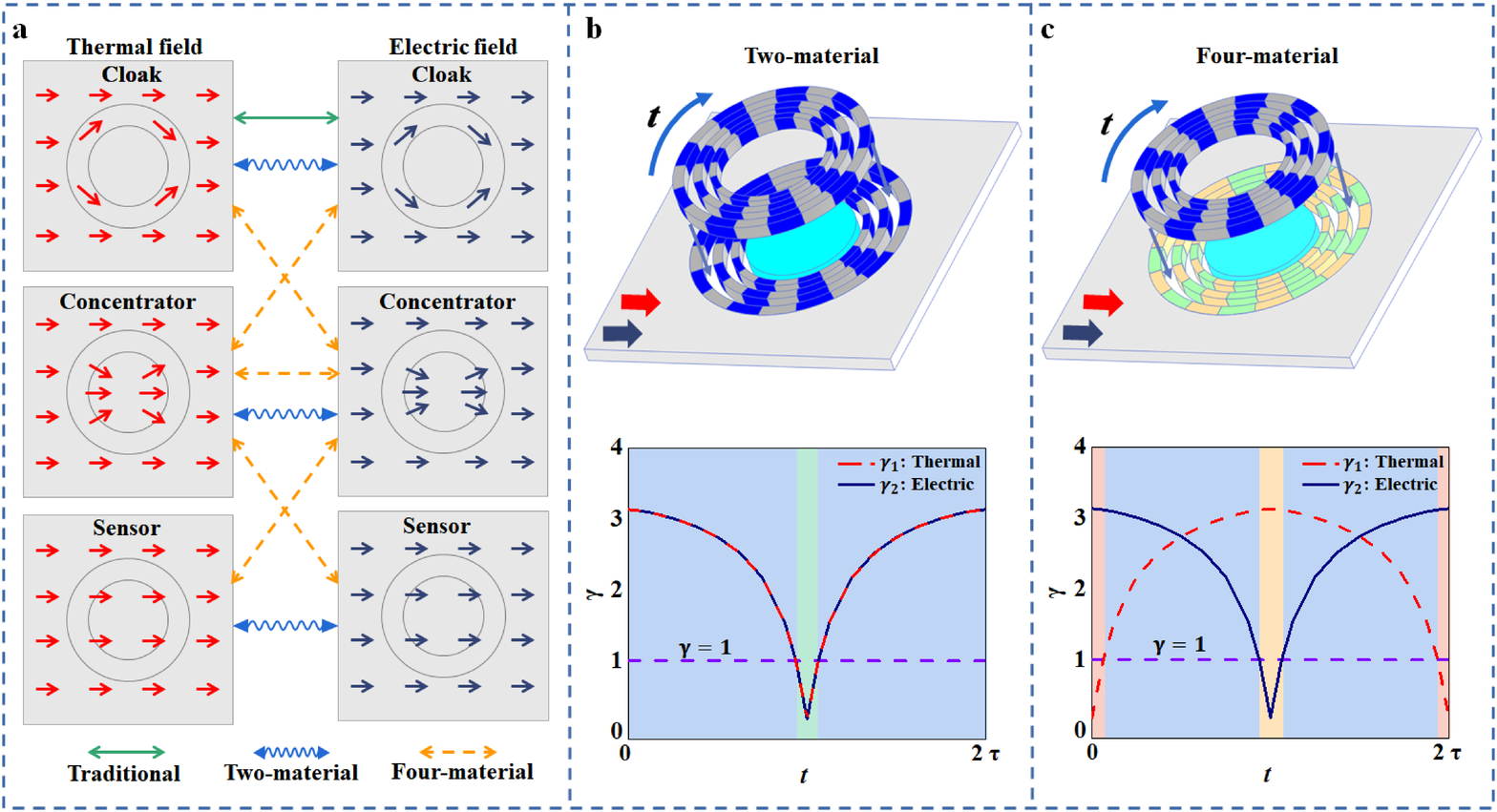}\\
	\caption{Concept of spatiotemporal multiphysics metamaterials. (a) Comparison between traditional and our proposed schemes. Traditional schemes only achieve one function combination, such as thermal cloaking plus electric cloaking. Our spatiotemporal multiphysics metamaterials comprising two (or four) materials realize three (or five) function combinations. (b) Two-material-based and (c) four-material-based checkerboard structures. The even-numbered layers rotate at a constant angular speed, changing the heat and electric currents in the central region and achieving continuous function adjustment. $2\tau$ is a rotation period. The heat and electric currents in the background are unchanged. $\gamma$ represents the ratio of central flow density to background flow density. $\gamma_1$ and $\gamma_2$ represent thermal and electric fields, respectively. The heat flow (current) in the central area is larger than in the background area, namely $\gamma_1>1$ ($\gamma_2>1$), corresponding to thermal (electric) concentrating; if  $\gamma_1<1$ ($\gamma_2<1$), the function is thermal (electric) cloaking; if $\gamma_1=1$ ($\gamma_2=1$), the function is thermal (electric) sensing.}
	\label{fig:boat1}
\end{figure}

\section{Results and Discussion}
\subsection{Theoretical Prediction}
We aim to realize various function combinations for thermal and electric fields in a single device and achieve continuous adjustment. For example, thermal cloaking plus electric concentrating is one function combination. Spatiotemporal multiphysics metamaterials composed of rotatable checkerboard structures can accomplish this goal. As shown in Fig.~1b and~c, the checkerboard unit consists of two or four isotropic materials, determining the type and number of function combinations for thermal and electric fields. Continuous function adjustment is realized by rotating the even-numbered layers of the checkerboard structure. The geometry also determines which function combination occurs during the rotating process. In this way, we can manipulate thermal and electric fields by controlling the spatial distribution and rotation time of the checkerboard unit.

The effective thermal and electric conductivities of the checkerboard determine the thermal and electric currents in the central region. According to the Keller theorem~\cite{P1964,JAP1975,JMP1985,IJHMT1992} and effective medium theory~\cite{JAP2010,OE2012,PRL2012,SR2013}, the effective thermal and electrical conductivity of the checkerboard structure in polar coordinates satisfy
\begin{equation}\label{KE}
\kappa_r \kappa_\theta=\kappa_1\kappa_2, ~~~\sigma_r\sigma_\theta=\sigma_1\sigma_2,
\end{equation}
where $\kappa_r$ ($\sigma_r$) and $\kappa_{\theta}$ ($\sigma_{\theta}$) are the effective radial and tangential thermal (electric) conductivities, $\kappa_1$ ($\sigma_1$) and $\kappa_2$ ($\sigma_2$) are the thermal (electric) conductivities of the two units that make up the checkerboard structure. We suppose that the thermal (electrical) conductivity of the two units satisfies $\kappa_{1}\kappa_{2}=\kappa_{r}\kappa_{\theta}=\kappa_b^2$ ($\sigma_{1}\sigma_{2}=\sigma_{r}\sigma_{\theta}=\sigma_b^2$). The product of the effective radial and tangential thermal (electrical) conductivities of the checkerboard structure is equal to the square of the background thermal (electrical) conductivity, so the background heat (electric) flow is not disturbed. The value of $\kappa_{r}/\kappa_{\theta}$ ($\sigma_{r}/\sigma_{\theta}$) affects the heat (electric) flow distribution in the central area:\\
 (1) When $\kappa_{r}/\kappa_{\theta}<1$ ($\sigma_{r}/\sigma_{\theta}<1$), the heat (electric) flow bypasses the central area to achieve thermal (electric) cloaking. If $\kappa_{r}/\kappa_{\theta}$ ($\sigma_{r}/\sigma_{\theta}$) is zero, the cloaking effect is perfect.\\
 (2) When $\kappa_{r}/\kappa_{\theta}=1$ ($\sigma_{r}\sigma_{\theta}=1$), the heat (electric) flow keeps unchanged in the central area to achieve thermal (electric) sensing.\\
 (3) When $\kappa_{r}/\kappa_{\theta}>1$ ( $\sigma_{r}/\sigma_{\theta}>1$), the heat (electric) flow is concentrated in the central area to achieve the effect of thermal (electric) concentrating.\\
Therefore, spatiotemporal function prediction is mainly based on calculating the effective radial and tangential thermal (electric) conductivities of the checkerboard structure.

\begin{figure}
	\centering
	\includegraphics[width=\linewidth]{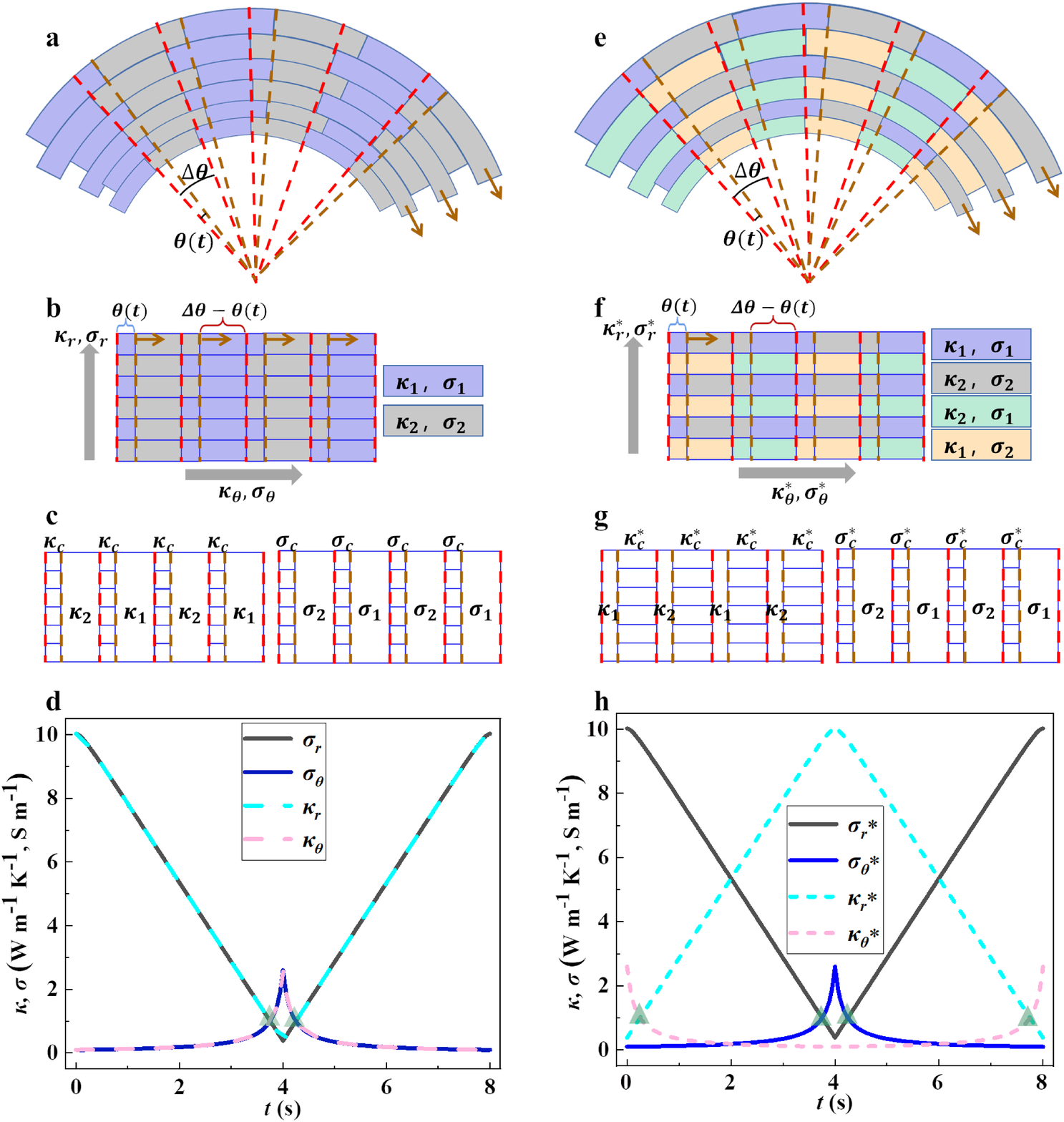}\\
	\caption{Schematic diagram of spatiotemporal multiphysics metamaterials. (a) Two-material-based checkerboard structure. The brown arrows represent the rotation direction of even layers. (b) The checkerboard structure stretched from (a), regarded as a multilayer structure. (c) Separate display of thermal and electric conductivities. (d) Theoretical results of the structure in (a). (e)-(h) Similar to (a)-(d), and the difference is that the checkerboard structure consists of four materials. The function marked by the green triangles is sensing.}
	\label{fig:boat1}
\end{figure}

We mainly consider two- and four-material-based checkerboard structures. As shown in Fig.~2a, we use two isotropic materials with constant thermal and electric conductivities to construct a checkerboard structure. The even layers rotate clockwise while the odd layers are fixed. The red dotted line in Fig.~2a does not move, and the brown dotted line rotates clockwise. $\Delta \theta$ is the central angle of a unit phase, $\tau$ is the time for the even-numbered layers to rotate $\Delta \theta$, and the rotation angle can be written as $\theta (t)=\Delta\theta t/\tau$. The position distribution of the checkerboard structure units returns to the initial one when the rotation time is $2\tau$. Thus, $2\tau$ is a rotation period and we only consider the rotating process in one period due to the periodicity of the checkerboard structure. When tuning the angle $\theta (t)$, we can divide the sector area corresponding to a central angle $\Delta \theta$ into homogeneous and staggered phases. The homogeneous phase is composed of the same material units, and its thermal and electric conductivities are $(\kappa_1, \sigma_1)$ or $(\kappa_2, \sigma_2)$. The staggered phase consists of two isotropic materials units interlaced with thermal and electric conductivities of $(\kappa_1, \sigma_1)$ and $(\kappa_2, \sigma_2)$, respectively. During a period $2\tau$, the central angle corresponding to staggered phases is
\begin{equation}\label{RA1}
 	\theta_{s}(t)=\left\lbrace 
 	\begin{array}{cc}
 		\frac{\Delta\theta}{\tau}t,  & 0<t<\tau\\
 		2\Delta\theta-\frac{\Delta\theta}{\tau}t.  & \tau<t<2\tau
 	\end{array}
 	\right.
\end{equation}
The angle $\theta_{s}(t)$ increases during $0<t<\tau$ and decreases during $\tau<t<2\tau$. The checkerboard structure returns to the original one at $2\tau$, with no staggering phase. The effective thermal conductivity of the staggered phases is determined by the material and shape parameters of the two units~\cite{PRAP2019}. The shape parameter of the staggered phases is a function of time and can be written as $\eta(t)=\ln (r_{i+1}/r_i)/\theta_{s}(t)$, where $r_i$ is the radius of the $i$-th layer. We replace $\eta$ in Equation (9) with $\eta(t)$ in Ref.~\cite{PRAP2019} to obtain the effective thermal conductivity $\kappa_c$ of the staggered phases; see Appendix A for detailed derivation. In this way, the entire checkerboard structure can be seen as three homogeneous materials interleaved, and the thermal conductivity is $\kappa_1$, $\kappa_c$, and $\kappa_2$. Note that the thermal conductivity of the staggered phase $\kappa_c$ is not the same from the tangential and radial directions.

We deform the fan-shaped structure in Fig.~2a into a multi-layer structure in Fig.~2b composed of staggered and homogeneous phases. The lateral thermal conductivity of the multi-layer structure is the tangential thermal conductivity of the fan-shaped structure, and the longitudinal thermal conductivity of the multi-layer structure is the radial thermal conductivity of the fan-shaped structure. The multi-layer structure can be approximated step-by-step using the effective medium theory to obtain the effective radial and tangential thermal conductivities of the rotatable checkerboard structure (see Appendix B for detailed derivation). The electric (thermal) conductivity of the checkerboard structure controls the electric (thermal) flow independently. As shown in Fig.~2c, we can thus separately consider the thermal and electric conductivities in the rotatable checkerboard structure. The thermal and electric conductivities of the two-material checkerboard structure satisfy the same variation. Therefore, we only need to use thermal conductivity as an example for theoretical calculations and finally replace $\kappa_1$ and $\kappa_2$ with $\sigma_1$ and $\sigma_2$ to get the effective electric conductivity. 

For clarity, we demonstrate the effective thermal and electric conductivity variation of the rotatable checkerboard structure during one period $2\tau$ in Fig.~2d with $\tau=4$~s. The parameters of the two materials that make up the checkerboard structure are ($\kappa_1=20~\rm {W m^{-1} K^{-1}}$, $\sigma_1=20~\rm{S m^{-1}}$) and ($\kappa_2=0.05~\rm {W m^{-1} K^{-1}}$, $\sigma_2=0.05~\rm{S m^{-1}}$). The thermal and electric conductivities of the background and central regions are $\kappa_b=1~\rm {W m^{-1} K^{-1}}$ and $\sigma_b=1~\rm{S m^{-1}}$. We bring these parameters into Equations~(\ref{R1}) and~(\ref{T1}) of the Appendix to get the results in Fig.~2d. The dashed and solid lines represent thermal and electric conductivities, respectively. The solid and dashed lines overlap, indicating that the thermal and electric conductivities satisfy the same equation. At $t=3.75$~s and $t=4.25$~s, the effective radial and tangential thermal (electric) conductivities are identical, indicating the thermal (electric) sensing function. The radial thermal (electric) conductivity is larger than the tangential thermal (electric) conductivity at $0<t<3.75$~s and $4.25<t<8$~s, namely $\kappa_r/\kappa_\theta >1$ ($\sigma_r/\sigma_\theta >1$), showing the thermal  (electric) concentrating function. The radial thermal (electric) conductivity is smaller than the tangential thermal (electric) conductivity at $3.75<t<4.25$~s, i.e., $\kappa_r/\kappa_\theta <1$ ($\sigma_r/\sigma_\theta <1$), demonstrating the thermal (electric) cloaking function. The thermal and electric fields of the two-material-based checkerboard structures have the same function, which can be regulated continuously over time. Thus, the spatiotemporal multiphysics metamaterials composed of two materials realize three function combinations.

Besides, we can also make thermal and electric functions switch with different rules. As shown in Fig.~2e, we design a checkerboard structure composed of four isotropic materials. The parameters of these four materials are $(\kappa_1,\sigma_1)$, $(\kappa_1,\sigma_2)$, $(\kappa_2,\sigma_1)$, and $(\kappa_2,\sigma_2)$. From the structures shown in Fig.~2c and~g, the electric conductivities of the checkerboard structure composed of two and four isotropic materials satisfy the same change. Therefore, the effective electric conductivities of the checkerboard structures of these two configurations are identical. In contrast, the thermal conductivity of the four-material-based checkerboard structure is different from that of the two-material-based checkerboard structure. The center angle corresponding to the staggered phase is
\begin{equation}\label{RA2}
	\theta_{s}^*(t)=\left\lbrace 
	\begin{array}{cc}
		\Delta\theta-\frac{\Delta\theta}{\tau}t,  & 0<t<\tau\\
		\frac{\Delta\theta}{\tau}t-\Delta\theta.  & \tau<t<2\tau
	\end{array}
	\right.
\end{equation}	
We replace all $\theta_{s}(t)$ in the thermal conductivity calculation of the two-material-based checkerboard structure with $\theta_{s}^*(t)$. Then, the effective thermal conductivity of the four-material-based checkerboard structure can be obtained (see Appendix C for detailed derivation).

As shown in Fig.~2h, we plot the effective thermal and electric conductivities as a function of time for the four-material-based checkerboard structure. The parameters of these four materials are $(\kappa_1,\sigma_1)$, $(\kappa_1,\sigma_2)$, $(\kappa_2,\sigma_1)$, and $(\kappa_2,\sigma_2)$, with $\kappa_1=20~\rm {Wm^{-1}K^{-1}}$, $\sigma_1=20 ~ \rm{Sm^{-1}}$, $\kappa_2=0.05~\rm {W m^{-1}K^{-1}}$, and $\sigma_2=0.05~\rm{Sm^{-1}}$. The parameters are substituted into Equations~(\ref{T1}) and~(\ref{CR2}) of the Appendix to get Fig.~2h. We theoretically predict that thermal sensing plus electric concentrating can be achieved at $t=0.25$~s and $t=7.75$~s. Thermal concentrating plus electric sensing can be obtained when $t=3.75$~ and $t=4.25$~s. The four-material-based checkerboard structure enables thermal cloaking plus electric concentrating before $0.25$~s and after $7.75$~s. Thermal concentrating plus electric cloaking are realized at $3.75<t<4.25$~s. The function combination at the rest time is thermal concentrating plus electric concentrating. Therefore, spatiotemporal multiphysics metamaterials made of four materials can realize five function combinations.

\subsection{Simulation Verification}
We perform finite-element simulations to verify the functionality of spatiotemporal multiphysics metamaterials. We first consider a rotatable checkerboard structure composed of two isotropic materials and take a hollow cylinder consisting of 15 sub-layers. Each layer is divided into 24 fan-shaped unit cells alternately arranged by two isotropic materials. The even layers of this hollow cylinder rotate clockwise with time, corresponding to Fig.~2a. The radius $r_i$ of each layer can be calculated by the formula $\ln(r_{i+1}/r_i) =\eta\Delta\theta$, with shape parameter $\eta=1/\sqrt{10}$, inner diameter $r_1=4$ cm, and unit center angle $\Delta\theta=2 \Pi/24$. We embed the checkerboard structure into a square with a side length of 45~cm. The thermal impedance between the checkerboard structural units is ignored. Temperature and potential differences are applied at the left and right boundaries to induce heat and electric currents through the entire structure. The left boundary temperature and potential are set at 373~K and 10~mV, and the right boundary is set at 273~K and 0~mV. The thermal and electric conductivities of the entire spatiotemporal metamaterial are consistent with the theory. The period is $2\tau=8$ s.

\begin{figure}
	\centering
	\includegraphics[width=\linewidth]{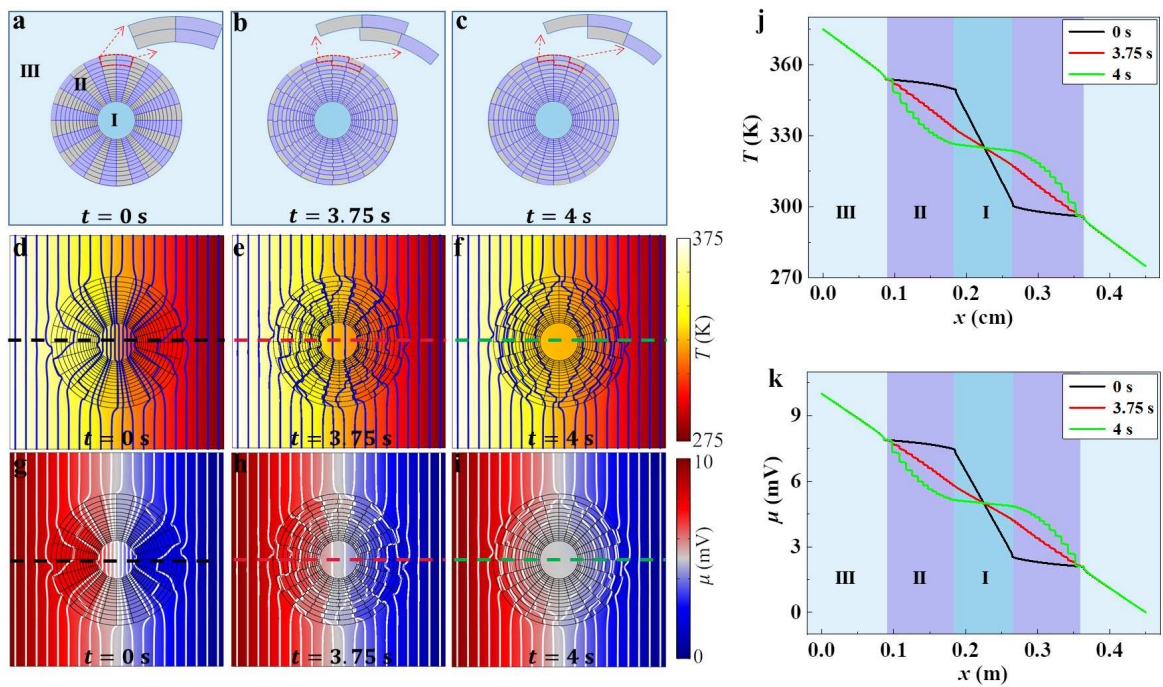}\\
	\caption{Schematic diagrams and simulation results of a two-material-based checkerboard structure. (a)-(c) Structures with $t=0$~s, $t=3.75$~s, and $t=4$~s. I, II, and III represent the central, checkerboard, and background areas. (d)-(f) Simulation results of the thermal field. Blue lines represent isotherms. (g)-(i) Simulation results of the electric field. White lines represent equipotential lines. (j) and (k) Data of the horizontal centerline from the simulation results in (d)-(f) and (g)-(i). The three temperature (or potential) lines in region III coincide.}
	\label{fig:boat1}
\end{figure}

The proposed theory can predict the functions for thermal and electric fields at any time. Spatiotemporal multiphysics metamaterials composed of two materials can realize three function combinations. As shown in Fig.~3, we select three moments corresponding to the chessboard structure of three function combinations in simulations, i.e., $t=0~\rm s$, $t=3.75~\rm s$, and $t=4~\rm s$. In the theoretical diagram Fig.~2d, these three moments correspond to concentrating, sensing, and cloaking. The background isotherms and equipotential lines remain unchanged at any time, indicating that the background heat and electric currents are not disturbed. At the initial time, the even-numbered layers have not yet started to rotate, and the checkerboard structure is composed of two uniform fan-shaped structures interlaced. The simulation results demonstrate the effect of thermal and electric concentrating. The heat and electric currents are concentrated in the central region without disturbing the background fields. In Fig.~3j and~k, the temperature and electric potential gradients in the central region are more significant than in the background, proving the function combination of thermal plus electric concentrating. At $t=3.75~\rm s$, the isotherm and equipotential line spacing in the central area is consistent with the background. In Fig.~3j and~k, the temperature and potential gradients in the central region are the same as the background, corresponding to thermal sensing plus electric sensing. The entire checkerboard structure consists of two materials interleaved at $t=4~\rm s$. The simulation results show that neither the isotherm nor the equipotential line enters the central region. The temperature and potential gradients in the central region are close to zero. Heat and electric currents do not enter the central area, so the effect of thermal cloaking plus electric cloaking is achieved. These three function combinations can also be achieved from $4~\rm s$ to $8~\rm s$. However, the function switch direction is opposite from $0~\rm s$ to $4~\rm s$ (see Appendix D for simulation results). Within a whole period of $8~\rm s$, the heat and electric flows in the central region of the checkerboard structure gradually decrease from a value higher than the background to zero and then gradually increase back to the initial state. We utilize a rotatable checkerboard structure to achieve time-responsive continuous adjustment of the thermal and electric functions. Thus, we verify the reliability of the theory with simulations and demonstrate that spatiotemporal multiphysics metamaterials composed of two materials can achieve three function combinations.

\begin{figure}
	\centering
	\includegraphics[width=\linewidth]{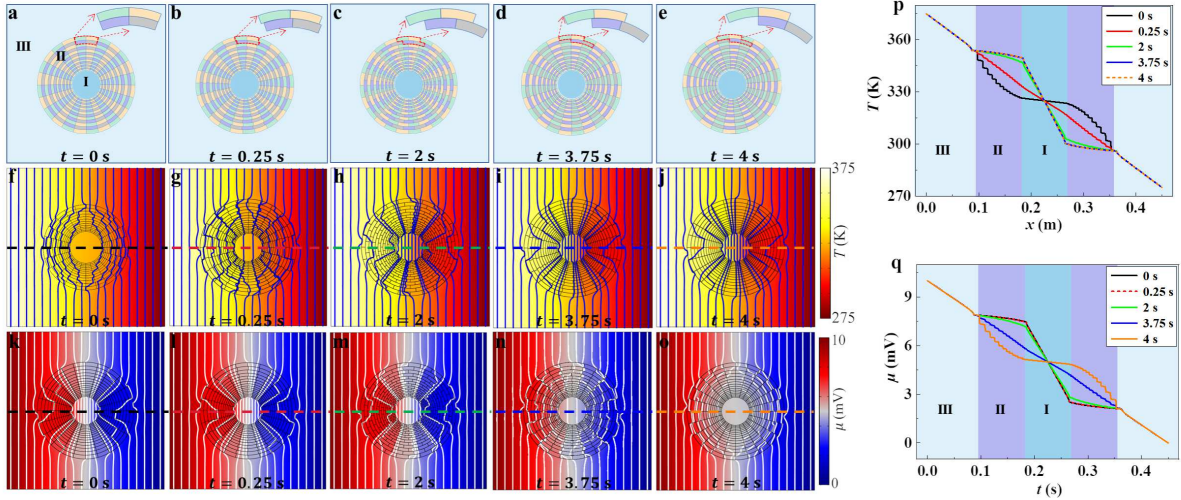}\\
	\caption{Schematic diagrams and simulation results of a four-material-based checkerboard structure. (a)-(e) Structures with $t=0~\rm s$, $t=0.25~\rm s$, $t=2~\rm s$, $t=3.75~\rm s$, and $t=4~\rm s$. (f)-(j) Simulation results of the thermal field. (k)-(o) Simulation results of the electric field. (p) and (q) Data of the horizontal centerline from the simulation results in (f)-(j) and (k)-(o). The five temperature (or potential) lines in region III coincide.}
	\label{fig:boat1}
\end{figure}

Thermal and electric fields can also perform different functions. We design a checkerboard structure composed of four isotropic materials. The parameters of the four isotropic materials are consistent with those in Fig.~2h. We show simulation results at five typical times in Fig.~4, namely $0~\rm s$, $0.25~\rm s$, $2~\rm s$, $3.75~\rm s$, and $4~\rm s$. The left boundary temperature and potential are set at 373~K and 10~mV, and the right boundary is set at 273~K and 0~mV. Neither heat nor electric flow in the background region is disturbed at any time. The heat flow bypasses the central area, and the current concentrates at $t=0~\rm s$, corresponding to thermal cloaking plus electric concentrating. The current bypasses the central region, and the heat flow focuses at $t=4~\rm s$, corresponding to electric cloaking plus thermal concentrating. We achieve the function combination of thermal sensing plus electric concentrating at $t=0.25~\rm s$ and thermal concentrating plus electric sensing at $t=3.75~\rm s$. The simulation results confirm that the theoretical predictions are indeed achieved. A checkerboard structure composed of four materials can regulate thermal and electric fields with different functions. At $t=2~\rm s$, the even-numbered layers in the checkerboard structure turn half the center corner of the unit. At this time, the radial thermal (electric) conductivity is higher than the tangential one, as shown in Fig.~2h. Both current and heat flows are concentrated in the central region, corresponding to thermal concentrating plus electric concentrating. According to Fig.~2h, thermal and electric concentrating is achieved from $0.25~\rm s$ to $3.75~\rm s$. Thus, spatiotemporal multiphysics metamaterials composed of four materials achieve five function combinations. From $4~\rm s$ to $8~\rm s$, these five function combinations can also be implemented (see Appendix E for simulation results). Our design realizes multi-field and multi-function operation with only the requirement of time control, which greatly weakens the limitation of fixed functions.

Spatiotemporal multiphysics metamaterials can also control thermal and electric functions by changing the geometry of the checkerboard structure unit. The geometric parameters change the number of function combinations of thermal and electric fields. For example, we consider a checkerboard structure composed of two isotropic materials and use the shape parameters $\eta=1$ and $\eta=1.5$ for simulation (see Appendix F for simulation results). The shape parameter $\eta=[\ln(r_{i+1}/r_i)]/(\Delta\theta)$ is only related to the unit shape and independent of time. A rotatable checkerboard structure with a shape parameter $\eta=1.5$ outputs only one function combination, i.e., thermal concentrating plus electric concentrating. The rotatable checkerboard structure with a shape parameter $\eta=1$ combines thermal sensing and electric sensing at $t=4~\rm s$. The rest time is the function combination of thermal and electric concentrating. The difference between these two structures is that the concentrating degree of the first type drops to zero, but the second type does not. These two structures can repeatedly realize the process of concentration degree from high to low and back to high for a long time, thereby contributing to timed heating and insulation. The object can be heated repeatedly by setting the period to keep it at a higher temperature. Similarly, the design can focus the current to maintain a high current density in a specific area. Thus, spatiotemporal multiphysics metamaterials can also flexibly change the function combinations via different structure shapes.

\section{Conclusions}
In summary, we propose the spatiotemporal multiphysics metamaterial with a rotatable checkerboard structure to simultaneously regulate thermal and electric fields. The introduced temporal dimension, i.e., the rotation time, can continuously control the effective thermal and electric conductivities, yielding the function switching of thermal and electric fields. The spatiotemporal checkerboard structure comprising two (or four) materials can realize three (or five) function combinations, demonstrating incredible performance. Based on the time-dependent effective medium theory, we predict the real-time thermal and electric functions, which are further confirmed by simulations. Our design is robust because only isotropic materials are required to achieve multiple function combinations. The method could also be employed in other multiphysics fields, such as electromagnetic-acoustic and thermal-magnetic fields.

\medskip
\noindent\textbf{Acknowledgements} \par 
\noindent We acknowledge financial support from the National Natural Science Foundation of China under Grant No.~12035004 and from the Science and Technology Commission of Shanghai Municipality under Grant No.~20JC1414700.

\medskip
\noindent\textbf{Conflict of Interest} \par 
\noindent The authors declare no conflicts of interest.

\medskip
\noindent\textbf{Data Availability Statement} \par
\noindent The data that support the findings of this study are available from the corresponding author upon reasonable request.

\clearpage
\newpage
\setcounter{equation}{0}
\renewcommand{\theequation}{A.\arabic{equation}}
\setcounter{figure}{0}
\renewcommand{\thefigure}{A.\arabic{figure}}

\textbf{Appendix A. Calculation of the effective thermal conductivity of the staggered phases of two-material-based checkerboard structures}

The effective thermal conductivity of a checkerboard structure consisting entirely of staggered phases (i.e., the rotating checkerboard structure at $\theta(t)=\Delta \theta$) is~\cite{PRAP2019}
\begin{equation}\label{SP}
	\begin{array}{c}
		\kappa_{cr}=\left\lbrace 
		\begin{array}{cc}
			\sqrt{\kappa_1 \kappa_2}\left[\left(1-\frac{2\sqrt{\kappa_1 \kappa_2}}{\kappa_1+\kappa_2}\right)(\eta-1)+1\right],  &\eta \le 1\\[2ex]
			\frac{\sqrt{\kappa_1 \kappa_2}}{\left(1-\frac{2\sqrt{\kappa_1 \kappa_2}}{\kappa_1+\kappa_2}\right)\left(\frac{1}{\eta}-1\right)+1},  &\eta \ge 1
		\end{array}
		\right.
		\\
		\\
		\kappa_{c\theta}=\left\lbrace 
		\begin{array}{cc}
			\frac{\sqrt{\kappa_1 \kappa_2}}{\left(1-\frac{2\sqrt{\kappa_1 \kappa_2}}{\kappa_1+\kappa_2}\right)(\eta-1)+1},  &\eta \le 1 \\[2ex]
			\sqrt{\kappa_1 \kappa_2}\left[\left(1-\frac{2\sqrt{\kappa_1 \kappa_2}}{\kappa_1+\kappa_2}\right)\left(\frac{1}{\eta}-1\right)+1\right],  &\eta \ge 1
		\end{array}
		\right.
	\end{array}
\end{equation}
where shape parameter $\eta=\ln (r_{i+1}/r_i)/\Delta\theta$ is only related to the geometry of the checkerboard structure, independent of time. We replace $\Delta \theta$ with $\theta_s(t)$ to get the shape parameter of the staggered phase over time, which can be written as 
\begin{equation}\label{RA1}
	\eta(t)=\ln (r_{i+1}/r_i)/\theta_s(t)=\left\lbrace 
	\begin{array}{cc}
		\eta\frac{\tau}{t},  & 0<t<\tau\\
		\eta\frac{\tau}{2\tau-t}\,  & \tau<t<2\tau
	\end{array}
	\right.
\end{equation}
The shape parameter here is not only related to the unit shape of the checkerboard but also related to time. The new shape parameter is taken into Equation~(\ref{SP}) to get the effective thermal conductivity $\kappa_c$ of staggered phases. In polar coordinates, $\kappa_c$ can be divided into radial thermal conductivity $\kappa_{cr}$ and tangential thermal conductivity $\kappa_{c\theta}$, respectively, 
\begin{equation}\label{CR1}
	\begin{array}{c}
		\kappa_{cr}=\left\lbrace 
		\begin{array}{cc}
			\frac{\sqrt{\kappa_1 \kappa_2}}{(1-\frac{2\sqrt{\kappa_1 \kappa_2}}{\kappa_1+\kappa_2})(\frac{t}{\eta \tau}-1)+1},  &0\le t \le \eta \tau
			\\[2ex]
			\sqrt{\kappa_1 \kappa_2}\left[(1-\frac{2\sqrt{\kappa_1 \kappa_2}}{\kappa_1+\kappa_2})(\frac{\eta \tau}{t}-1)+1\right],  & \eta \tau \le t \le  \tau\\[2ex]
			
			\sqrt{\kappa_1 \kappa_2}\left[(1-\frac{2\sqrt{\kappa_1 \kappa_2}}{\kappa_1+\kappa_2})(\frac{\eta \tau}{2\tau-t}-1)+1\right],
			&\tau \le t \le (2-\eta) \tau\\[2ex]
			
			\frac{\sqrt{\kappa_1 \kappa_2}}{(1-\frac{2\sqrt{\kappa_1 \kappa_2}}{\kappa_1+\kappa_2})(\frac{2\tau-t}{\eta \tau}-1)+1},   & (2-\eta) \tau \le t \le 2\tau \\[2ex]	
		\end{array}
		\right.
		\\
		\\
		\kappa_{c\theta}=\left\lbrace 
		\begin{array}{cc}
			\sqrt{\kappa_1 \kappa_2}\left[(1-\frac{2\sqrt{\kappa_1 \kappa_2}}{\kappa_1+\kappa_2})(\frac{t}{\eta \tau}-1)+1\right], &0 \le t \le \eta \tau\\[2ex]
			
			\frac{\sqrt{\kappa_1 \kappa_2}}{(1-\frac{2\sqrt{\kappa_1 \kappa_2}}{\kappa_1+\kappa_2})(\frac{\eta \tau}{t}-1)+1},  & \eta \tau \le t \le \tau\\[2ex]
			
			\frac{\sqrt{\kappa_1 \kappa_2}}{(1-\frac{2\sqrt{\kappa_1 \kappa_2}}{\kappa_1+\kappa_2})(\frac{\eta \tau}{2\tau-t}-1)+1},  & \tau \le t \le (2-\eta) \tau \\[2ex]
			
			\sqrt{\kappa_1 \kappa_2}\left[(1-\frac{2\sqrt{\kappa_1 \kappa_2}}{\kappa_1+\kappa_2})(\frac{2\tau-t}{\eta \tau}-1)+1\right]. & (2-\eta) \tau \le t \le 2\tau
		\end{array}
		\right.
	\end{array}
\end{equation}
The effective radial and tangential thermal conductivity of the staggered phase over the entire period $2\tau$ is composed of segmentation functions, which is a function of time. The effective thermal conductivity of the interleaved phase is also related to the shape parameter $\eta$ of the checkerboard structure unit, and Equation~(\ref{CR1}) gives the result of $\eta \le 1$. If $\eta>1$, it is easy to remove the piecewise function with a contradictory time range in Equation~(\ref{CR1}).

\textbf{Appendix B. Calculation of the effective thermal conductivity of two-material-based checkerboard structures}

\begin{figure}
	\centering
	\includegraphics[width=\linewidth]{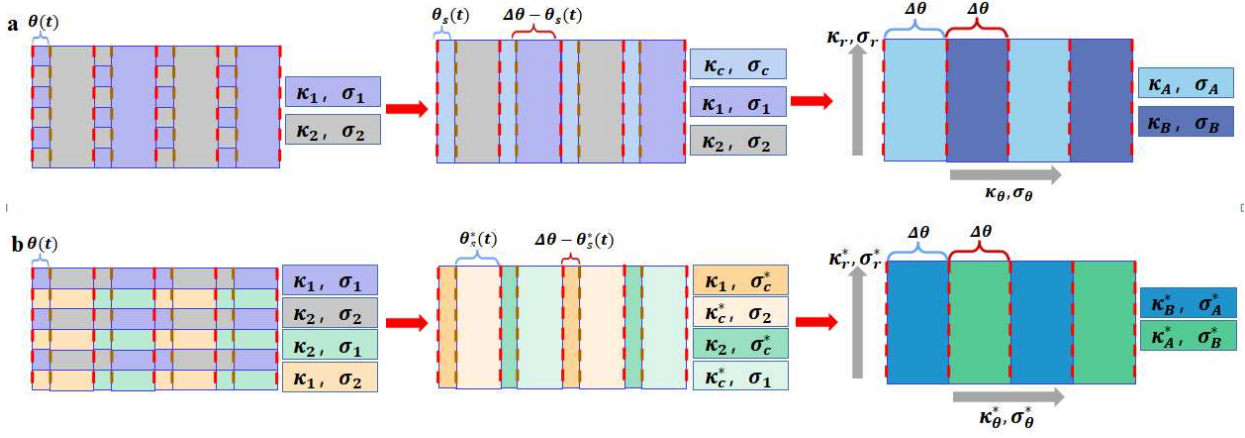}\\
	\caption{Evolution of rotatable checkerboard structures based on the effective medium theory. The rotatable checkerboard structure consists of (a) two materials and (b) four materials.}
	\label{fig:boat1}
\end{figure}

The fan-shaped structure of the checkerboard structure is morphed into a multi-layered structure composed of staggered and homogeneous phases shown in Fig. A.1a. We can combine a staggered and homogeneous phase to form a unit. The thermal conductivity of the staggered phase is $\kappa_c$, and the thermal conductivity of the homogeneous phase can be $\kappa_1$ or $\kappa_2$. Therefore, we can obtain two kinds of units: the combination of $\kappa_c$ and $\kappa_1$ and the combination of $\kappa_c$ and $\kappa_2$. According to the effective medium theory, the radial and tangential effective thermal conductivity of these two units can be calculated by
\begin{equation}\label{AB1}
	\begin{array}{c}
		\kappa_{Ar}=\frac{\theta_s(t)}{\Delta\theta}\kappa_{cr}+\frac{\Delta\theta-\theta_s(t)}{\Delta\theta}\kappa_2,
		\kappa_{Br}=\frac{\theta_s(t)}{\Delta\theta}\kappa_{cr}+\frac{\Delta\theta-\theta_s(t)}{\Delta\theta}\kappa_1.
		\\
		\\
		\frac{1}{\kappa_{A\theta}}=\frac{\theta_s(t)}{\Delta\theta} \frac{1}{\kappa_{c\theta}}+\frac{\Delta\theta-\theta_s(t)}{\Delta\theta}\frac{1}{\kappa_2},
		\frac{1}{\kappa_{B\theta}}=\frac{\theta_s(t)}{\Delta\theta}\frac{1}{\kappa_{c\theta}}+\frac{\Delta\theta-\theta_s(t)}{\Delta\theta}\frac{1}{\kappa_1}.
	\end{array}
\end{equation}
where $\theta_s(t)/\Delta\theta$ is the proportion of staggered phase in a unit, and $(\Delta\theta-\theta_s(t))/\Delta\theta$ corresponds to homogeneous phase. Then the whole structure can be regarded as a multi-layer structure composed of two kinds of units staggered. Using the effective medium theory again, 
\begin{equation}\label{RT1}
	\begin{array}{c}
		\kappa_r=\frac{\kappa_{Ar}+\kappa_{Br}}{2}
		\\
		\frac{1}{\kappa_\theta}=\frac{1}{2\kappa_{A\theta}}+\frac{1}{2\kappa_{B\theta}}.
	\end{array}
\end{equation}
The corresponding central angles of two units A and B in the chessboard structure are the same. The radial and tangential effective thermal conductivities of the rotatable checkerboard structure composed of two isotropic materials can be obtained by solving Equation~(\ref{CR1})-Equation~(\ref{RT1}), which can be written as
\begin{equation}\label{R1}
	\begin{array}{cc}
		\kappa_{r}=\left\lbrace 
		\begin{array}{cc}
			\frac{t}{\tau}\frac{\sqrt{\kappa_1 \kappa_2}}{(1-\frac{2\sqrt{\kappa_1 \kappa_2}}{\kappa_1+\kappa_2})(\frac{t}{\eta \tau}-1)+1}+(\frac{1}{2}-\frac{t}{2\tau})(\kappa_1+\kappa_2),  &0 \le t \le \eta \tau \\[2ex] 
			
			\frac{t}{\tau}\sqrt{\kappa_1 \kappa_2}\left[(1-\frac{2\sqrt{\kappa_1 \kappa_2}}{\kappa_1+\kappa_2})(\eta\frac{\tau}{t}-1)+1\right]+(\frac{1}{2}-\frac{t}{2\tau})(\kappa_1+\kappa_2),  &\eta \tau \le t \le  \tau \\[2ex]
			
			(2-\frac{t}{\tau})	
			\sqrt{\kappa_1 \kappa_2}\left[(1-\frac{2\sqrt{\kappa_1 \kappa_2}}{\kappa_1+\kappa_2})(\frac{\eta \tau}{2\tau-t}-1)+1\right]+(\frac{t}{2\tau}-\frac{1}{2})(\kappa_1+\kappa_2), &\tau \le t \le (2-\eta) \tau\\[2ex]
			
			(2-\frac{t}{\tau})	
			\frac{\sqrt{\kappa_1 \kappa_2}}{(1-\frac{2\sqrt{\kappa_1 \kappa_2}}{\kappa_1+\kappa_2})(\frac{2\tau-t}{\eta \tau}-1)+1}+(\frac{t}{2\tau}-\frac{1}{2})(\kappa_1+\kappa_2), &(2-\eta) \tau \le t \le 2\tau \\[2ex]
		\end{array}
		\right.
		\\
		\\
		\kappa_{\theta}=\left\lbrace 
		\begin{array}{cc}
			\frac{\kappa_1\kappa_2}{\frac{t}{\tau}\frac{\sqrt{\kappa_1 \kappa_2}}{(1-\frac{2\sqrt{\kappa_1 \kappa_2}}{\kappa_1+\kappa_2})(\frac{t}{\eta \tau}-1)+1}+(\frac{1}{2}-\frac{t}{2\tau})(\kappa_1+\kappa_2)},   &0 \le t \le \eta \tau \\[3ex]
			
			\frac{\kappa_1\kappa_2}{\sqrt{\kappa_1\kappa_2}\frac{t}{\tau}\left[(1-\frac{2\sqrt{\kappa_1\kappa_2}}{\kappa_1+\kappa_2})(\frac{\eta \tau}{t}-1)+1\right]+(\frac{1}{2}-\frac{t}{2\tau})(\kappa_1+\kappa_2)},  &\eta \tau \le t \le  \tau \\[3ex]
			
			\frac{\kappa_1\kappa_2}{(2-\frac{t}{\tau})	
				\sqrt{\kappa_1 \kappa_2}\left[(1-\frac{2\sqrt{\kappa_1 \kappa_2}}{\kappa_1+\kappa_2})(\frac{\eta \tau}{2\tau-t}-1)+1\right]+(\frac{t}{2\tau}-\frac{1}{2})(\kappa_1+\kappa_2)},   &\tau \le t \le (2-\eta) \tau\\[3ex] 
			
			\frac{\kappa_1\kappa_2}{(2-\frac{t}{\tau})	
				\frac{\sqrt{\kappa_1 \kappa_2}}{(1-\frac{2\sqrt{\kappa_1 \kappa_2}}{\kappa_1+\kappa_2})(\frac{2\tau-t}{\eta \tau}-1)+1}+(\frac{t}{2\tau}-\frac{1}{2})(\kappa_1+\kappa_2)}.   & (2-\eta) \tau \le t \le 2\tau		
		\end{array}
		\right.
	\end{array}
\end{equation}
Similarly, the radial and tangential effective electric conductivities of the rotatable checkerboard structure composed of two isotropic materials can be written as
\begin{equation}\label{T1}
	\begin{array}{cc}
		\sigma_{r}=\left\lbrace 
		\begin{array}{cc}
			\frac{t}{\tau}\frac{\sqrt{\sigma_1 \sigma_2}}{(1-\frac{2\sqrt{\sigma_1 \sigma_2}}{\sigma_1+\sigma_2})(\frac{t}{\eta \tau}-1)+1}+(\frac{1}{2}-\frac{t}{2\tau})(\sigma_1+\sigma_2),  &0 \le t \le \eta \tau \\[2ex]
			
			\frac{t}{\tau}\sqrt{\sigma_1 \sigma_2}\left[(1-\frac{2\sqrt{\sigma_1 \sigma_2}}{\sigma_1+\sigma_2})(\eta\frac{\tau}{t}-1)+1\right]+(\frac{1}{2}-\frac{t}{2\tau})(\sigma_1+\sigma_2),  & \eta \tau \le t \le \tau \\[2ex]
			
			(2-\frac{t}{\tau})	\sqrt{\sigma_1 \sigma_2}\left[(1-\frac{2\sqrt{\sigma_1 \sigma_2}}{\sigma_1+\sigma_2})(\eta\frac{\tau}{2\tau-t}-1)+1\right]+(\frac{t}{2\tau}-\frac{1}{2})(\sigma_1+\sigma_2),  &\tau \le t \le (2-\eta) \tau\\[2ex] 
			
			(2-\frac{t}{\tau})\frac{\sqrt{\sigma_1 \sigma_2}}{(1-\frac{2\sqrt{\sigma_1 \sigma_2}}{\sigma_1+\sigma_2})(\frac{2\tau-t}{\eta \tau}-1)+1}+(\frac{t}{2\tau}-\frac{1}{2})(\sigma_1+\sigma_2),   &(2-\eta) \tau \le t \le 2\tau	 \\[2ex]
		\end{array}
		\right.
		\\
		\\
		\sigma_{\theta}=\left\lbrace 
		\begin{array}{cc}
			\frac{\sigma_1\sigma_2}{\frac{t}{\tau}\frac{\sqrt{\sigma_1 \sigma_2}}{(1-\frac{2\sqrt{\sigma_1 \sigma_2}}{\sigma_1+\sigma_2})(\frac{t}{\eta \tau}-1)+1}+(\frac{1}{2}-\frac{t}{2\tau})(\sigma_1+\sigma_2)},  &0 \le t \le \eta \tau \\[3ex]
			
			\frac{\sigma_1\sigma_2}{\frac{t}{\tau}\sqrt{\sigma_1 \sigma_2}\left[(1-\frac{2\sqrt{\sigma_1 \sigma_2}}{\sigma_1+\sigma_2})(\eta\frac{\tau}{t}-1)+1\right]+(\frac{1}{2}-\frac{t}{2\tau})(\sigma_1+\sigma_2)},  & \eta \tau \le t \le \tau \\[3ex]
			
			\frac{\sigma_1\sigma_2}{(2-\frac{t}{\tau})	\sqrt{\sigma_1 \sigma_2}\left[(1-\frac{2\sqrt{\sigma_1 \sigma_2}}{\sigma_1+\sigma_2})(\eta\frac{\tau}{2\tau-t}-1)+1\right]+(\frac{t}{2\tau}-\frac{1}{2})(\sigma_1+\sigma_2)},  &\tau \le t \le (2-\eta) \tau\\[3ex] 
			
			\frac{\sigma_1\sigma_2}{(2-\frac{t}{\tau})\frac{\sqrt{\sigma_1 \sigma_2}}{(1-\frac{2\sqrt{\sigma_1 \sigma_2}}{\sigma_1+\sigma_2})(\frac{2\tau-t}{\eta \tau}-1)+1}+(\frac{t}{2\tau}-\frac{1}{2})(\sigma_1+\sigma_2)}.   &(2-\eta) \tau \le t \le 2\tau	
		\end{array}
		\right.
	\end{array}
\end{equation}	
Equation~(\ref{R1}) and Equation~(\ref{T1}) indicates that the effective thermal conductivity and electric conductivity satisfy the relation $\kappa_r \kappa_{\theta}=\kappa_1\kappa_2$ and $\sigma_r\sigma_{\theta}=\sigma_1\sigma_2$, which is consistent with Equation (1) in the main text.

\textbf{Appendix C. Calculation of the effective thermal and electric conductivities of four-material-based checkerboard structures}

The electric conductivity of the checkerboard structure composed of two isotropic materials and the checkerboard structure composed of four isotropic materials satisfy the same change, as shown in Fig. A.1a and b. Thus, we can get the relationship $\sigma_r^*=\sigma_r$ and $\sigma_{\theta}^*=\sigma_{\theta}$. Next, we calculate the effective thermal conductivity of the four-material checkerboard structure. When the even-numbered layers rotate through the angle $\theta(t)$, the central angle corresponding to the staggered area is $\theta_{s}^*(t)$ of the main text. The corresponding shape parameters of the staggering area of the checkerboard structure of the four materials are rewritten as
\begin{equation}\label{RA2}
	\eta^*(t)=\ln (r_{i+1}/r_i)/\theta_s^*(t)=\left\lbrace 
	\begin{array}{cc}
		\eta\frac{\tau}{\tau-t},  & 0<t<\tau\\
		\eta\frac{\tau}{t-\tau}\,  & \tau<t<2\tau
	\end{array}
	\right.
\end{equation}
The effective radial and tangential thermal conductivity of the interleaved phase obtained by bringing the new shape parameters Equation~(\ref{RA2}) into Equation~(\ref{SP}) can be written as
\begin{equation}\label{R2}
	\begin{array}{cc}
		\kappa_{cr}^*=\left\lbrace 
		\begin{array}{cc}
			\sqrt{\kappa_1 \kappa_2}\left[(1-\frac{2\sqrt{\kappa_1 \kappa_2}}{\kappa_1+\kappa_2})(\eta\frac{\tau}{\tau-t}-1)+1\right],  &0 \le t \le \tau(1-\eta)\\[2ex]
			
			\frac{\sqrt{\kappa_1 \kappa_2}}{(1-\frac{2\sqrt{\kappa_1 \kappa_2}}{\kappa_1+\kappa_2})(\frac{\tau-t}{\eta\tau}-1)+1},  & \tau(1-\eta) \le t \le \tau \\[2ex]
			
			\frac{\sqrt{\kappa_1 \kappa_2}}{(1-\frac{2\sqrt{\kappa_1 \kappa_2}}{\kappa_1+\kappa_2})(\frac{t-\tau}{\eta\tau}-1)+1}, &\tau \le t \le \tau(1+\eta) \\[2ex]
			
			\sqrt{\kappa_1 \kappa_2}\left[(1-\frac{2\sqrt{\kappa_1 \kappa_2}}{\kappa_1+\kappa_2})(\eta\frac{\tau}{t-\tau}-1)+1\right],  &\tau(1+\eta) \le t \le 2\tau\\[2ex]
		\end{array}
		\right.\\
		\kappa_{c\theta}^*=\left\lbrace 
		\begin{array}{cc}
			\frac{\sqrt{\kappa_1 \kappa_2}}{(1-\frac{2\sqrt{\kappa_1 \kappa_2}}{\kappa_1+\kappa_2})(\eta\frac{\tau}{\tau-t}-1)+1},   &0 \le t \le \tau(1-\eta)\\[2ex]
			
			\sqrt{\kappa_1 \kappa_2}\left[(1-\frac{2\sqrt{\kappa_1 \kappa_2}}{\kappa_1+\kappa_2})(\frac{\tau-t}{\eta\tau}-1)+1\right],   & \tau(1-\eta) \le t \le \tau \\[2ex]
			
			\sqrt{\kappa_1 \kappa_2}\left[(1-\frac{2\sqrt{\kappa_1 \kappa_2}}{\kappa_1+\kappa_2})(\frac{t-\tau}{\eta\tau}-1)+1\right],  &\tau \le t \le \tau(1+\eta) \\[2ex]
			
			\frac{\sqrt{\kappa_1 \kappa_2}}{(1-\frac{2\sqrt{\kappa_1 \kappa_2}}{\kappa_1+\kappa_2})(\eta\frac{\tau}{t-\tau}-1)+1}.  &\tau(1+\eta) \le t \le 2\tau\\[2ex]
		\end{array}
		\right.		
	\end{array}
\end{equation}
Similar to the checkerboard structure of the two materials, $\kappa_c^*$ is combined with  $\kappa_1$ and  $\kappa_2$, respectively. The entire checkerboard structure is regarded as the staggered composition of two elements A and B, whose thermal conductivity is written as
\begin{equation}\label{AB2}
	\begin{array}{c}
		\kappa_{Ar}^*=\frac{\theta_s^*(t)}{\Delta\theta}\kappa_{cr}^*+\frac{\Delta\theta-\theta_s^*(t)}{\Delta\theta}\kappa_2,
		\kappa_{Br}^*=\frac{\theta_s^*(t)}{\Delta\theta}\kappa_{cr}^*+\frac{\Delta\theta-\theta_s^*(t)}{\Delta\theta}\kappa_1.
		\\
		\\
		\frac{1}{\kappa_{A\theta}^*}=\frac{\theta_s^*(t)}{\Delta\theta} \frac{1}{\kappa_{c\theta}^*}+\frac{\Delta\theta-\theta_s^*(t)}{\Delta\theta}\frac{1}{\kappa_2},
		\frac{1}{\kappa_{B\theta}^*}=\frac{\theta_s^*(t)}{\Delta\theta}\frac{1}{\kappa_{c\theta}^*}+\frac{\Delta\theta-\theta_s^*(t)}{\Delta\theta}\frac{1}{\kappa_1}.
	\end{array}
\end{equation}
Again using the effective medium theory, the effective thermal conductivity of the entire checkerboard structure can be obtained,
\begin{equation}\label{RT2}
	\begin{array}{c}
		\kappa_r^*=\frac{\kappa_{Ar}^*+\kappa_{Br}^*}{2}
		\\
		\frac{1}{\kappa_\theta^*}=\frac{1}{2\kappa_{A\theta}^*}+\frac{1}{2\kappa_{B\theta}^*}.
	\end{array}
\end{equation}

After calculation, the effective thermal conductivity of the rotatable checkerboard structure composed of four isotropic materials can be written as    
\begin{equation}\label{CR2}
	\begin{array}{cc}
		\kappa_{r}^*=\left\lbrace 
		\begin{array}{cc}
			(1-\frac{t}{\tau})\sqrt{\kappa_1 \kappa_2}\left[(1-\frac{2\sqrt{\kappa_1 \kappa_2}}{\kappa_1+\kappa_2})(\frac{\eta \tau}{\tau-t}-1)+1\right]+\frac{t}{2\tau}(\kappa_1+\kappa_2),  &0 \le t \le \tau(1-\eta)\\[2ex]
			
			(1-\frac{t}{\tau})\frac{\sqrt{\kappa_1 \kappa_2}}{(1-\frac{2\sqrt{\kappa_1 \kappa_2}}{\kappa_1+\kappa_2})(\frac{\tau-t}{\eta \tau}-1)+1}+\frac{t}{2\tau}(\kappa_1+\kappa_2), &\tau(1-\eta) \le t \le \tau \\[2ex]
			
			(\frac{t}{\tau}-1)\frac{\sqrt{\kappa_1 \kappa_2}}{(1-\frac{2\sqrt{\kappa_1 \kappa_2}}{\kappa_1+\kappa_2})(\frac{t-\tau}{\eta\tau}-1)+1}+(1-\frac{t}{2\tau})(\kappa_1+\kappa_2), &\tau \le t \le \tau(1+\eta) \\[2ex]
			
			(\frac{t}{\tau}-1)\sqrt{\kappa_1 \kappa_2}\left[(1-\frac{2\sqrt{\kappa_1 \kappa_2}}{\kappa_1+\kappa_2})(\frac{\eta\tau}{t-\tau}-1)+1\right]+(1-\frac{t}{2\tau})(\kappa_1+\kappa_2),  &\tau(1+\eta) \le t \le 2\tau
		\end{array}
		\right.\\
		\\
		\kappa_{\theta}^*=\left\lbrace 
		\begin{array}{cc}
			\frac{\kappa_1 \kappa_2}{(1-\frac{t}{\tau})\sqrt{\kappa_1 \kappa_2}\left[(1-\frac{2\sqrt{\kappa_1 \kappa_2}}{\kappa_1+\kappa_2})(\frac{\eta \tau}{\tau-t}-1)+1\right]+\frac{t}{2\tau}(\kappa_1+\kappa_2)},  &0 \le t \le \tau(1-\eta)\\[3ex]
			
			\frac{\kappa_1 \kappa_2}{(1-\frac{t}{\tau})\frac{\sqrt{\kappa_1 \kappa_2}}{(1-\frac{2\sqrt{\kappa_1 \kappa_2}}{\kappa_1+\kappa_2})(\frac{\tau-t}{\eta \tau}-1)+1}+\frac{t}{2\tau}(\kappa_1+\kappa_2)}, &\tau(1-\eta) \le t \le \tau \\[3ex]
			
			\frac{\kappa_1 \kappa_2}{(\frac{t}{\tau}-1)\frac{\sqrt{\kappa_1 \kappa_2}}{(1-\frac{2\sqrt{\kappa_1 \kappa_2}}{\kappa_1+\kappa_2})(\frac{t-\tau}{\eta\tau}-1)+1}+(1-\frac{t}{2\tau})(\kappa_1+\kappa_2)}, &\tau \le t \le \tau(1+\eta) \\[3ex]
			
			\frac{\kappa_1 \kappa_2}{(\frac{t}{\tau}-1)\sqrt{\kappa_1 \kappa_2}\left[(1-\frac{2\sqrt{\kappa_1 \kappa_2}}{\kappa_1+\kappa_2})(\frac{\eta\tau}{t-\tau}-1)+1\right]+(1-\frac{t}{2\tau})(\kappa_1+\kappa_2)},  &\tau(1+\eta) \le t \le 2\tau
		\end{array}
		\right.		
	\end{array}
\end{equation}

\textbf{Appendix D. Simulation results of two-material-based checkerboard structures during $\tau \le t \le 2\tau$}

\begin{figure}
	\centering
	\includegraphics[width=\linewidth]{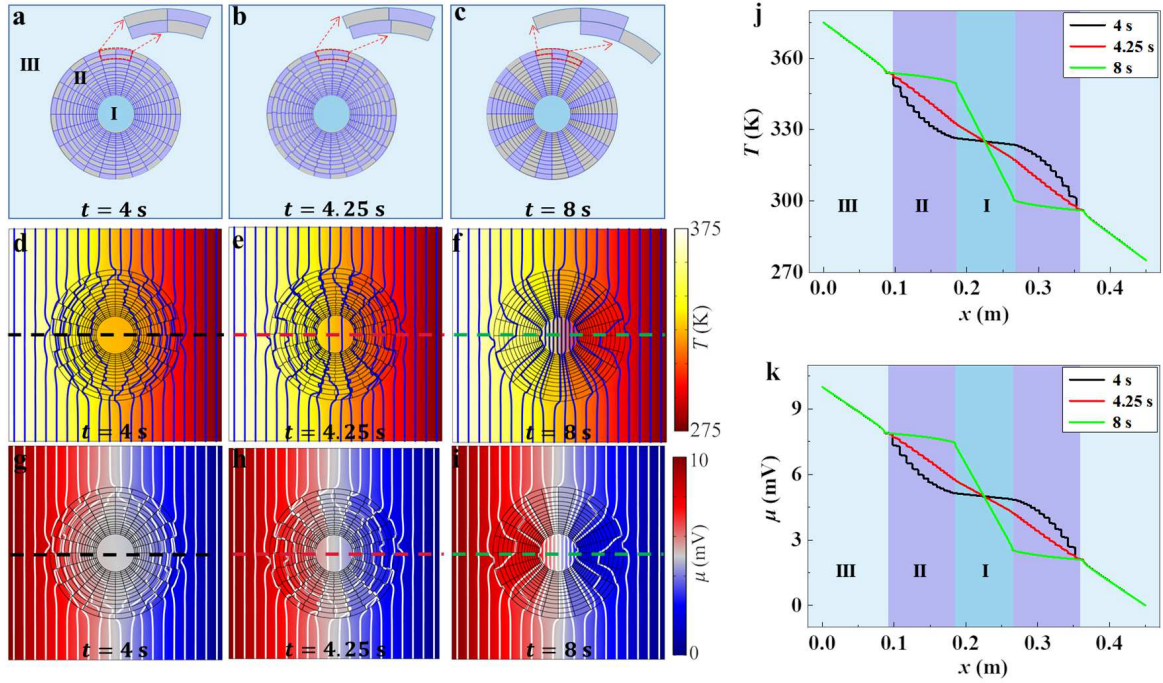}\\
	\caption{(a)-(c) Structures with $t=4~\rm s$, $t=4.25~\rm s$, $t=8~\rm s$. (d)-(f) and (g)-(i) Simulation results of the thermal and electric fields. (j) and (k) Data of the horizontal center line from the simulation results of (d)-(f) and (g)-(i). The three temperature (or potential) lines in region III coincide.}
	\label{fig:boat1}
\end{figure}

Two-material checkerboard structures appear in time 4 s $\le t \le $8 s three functional combinations. When $t=4~\rm s$, the function combination is thermal cloak plus electric cloak; when $t=4.25~\rm s$, the function combination is thermal sensor plus electric sensor; when $t=8~\rm s$, the function combination is thermal concentrator plus electric concentrator. Over time, the functions of the thermal and electric fields change from cloaks to sensors to concentrators. The direction of functional change in the second half cycle is the opposite of that in the first half cycle.

\textbf{Appendix E. Simulation results of four-material-based checkerboard structures during $\tau \le t \le 2\tau$}

The checkerboard structure of four materials completes five functional combinations in time 4 s $\le t \le $8 s. With the increase of time, the concentration degree of the thermal concentrator gradually decreases to zero at $t=7.75~\rm s$, and the function becomes a sensor. The heat flow in the middle region continues to decline, finally changing function to cloak at $t=8~\rm s$. The direction of the phase transition of the electric field is opposite to that of the thermal field. The five function combinations obtained at these five moments are in agreement with the results predicted by the theory.

\begin{figure}
	\centering
	\includegraphics[width=\linewidth]{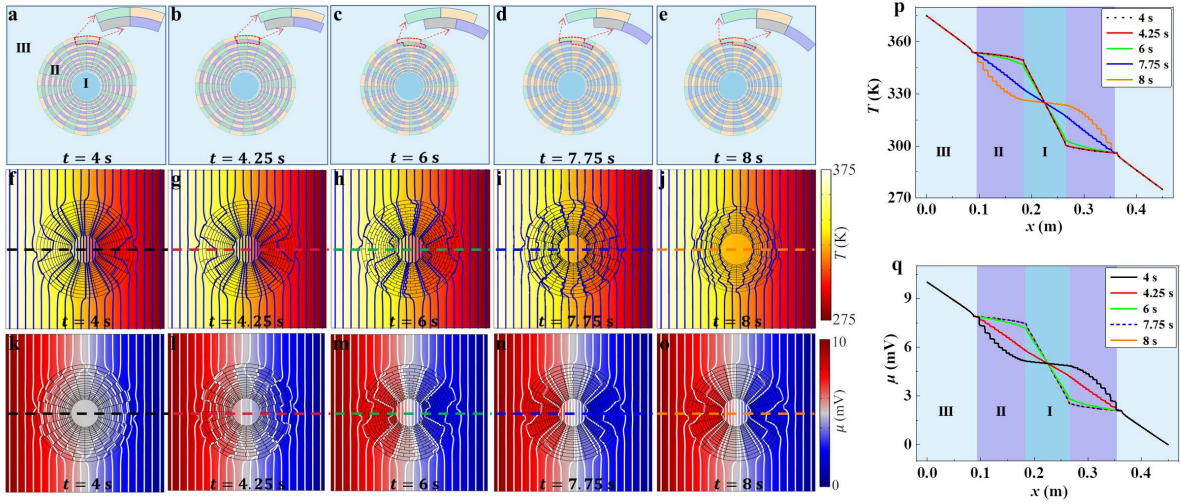}\\
	\caption{(a)-(e) Structures with $t=4~\rm s$, $t=4.25~\rm s$, $t=6~\rm s$, $t=7.75~\rm s$, $t=8~\rm s$. (f)-(j) and (k)-(o) Simulation results of the thermal and electric fields. (p) and (q) Data of the horizontal center line from the simulation results of (f)-(j) and (k)-(o). The five temperature (or potential) lines in region III coincide.}
	\label{fig:boat1}
\end{figure}

\textbf{Appendix F. Theory and simulation results of checkerboard structures with different shapes}

The checkerboard structure is composed of two isotropic materials. The parameters of the two isotropic materials are ($\kappa_1=10~ \rm {Wm^{-1}K^{-1}}$, $\sigma_1=10 ~ \rm{Sm^{-1}}$) and ($\kappa_2=0.1 ~\rm {Wm^{-1}K^{-1}}$, $\sigma_2=0.1~  \rm{Sm^{-1}}$), respectively. We use $\eta$ and $t$ in Equations~(\ref{R1}) (Equations~(\ref{T1})) as independent variables to compare the radial and tangential thermal (electric) conductivity. The rest of the parameters are the same as those in Fig. 3 of the main text. The phase diagram corresponding to the drawable function with $\eta$ as the horizontal axis and $t$ as the vertical axis is shown in Fig. A.4a, where $\tau=4~\rm s$ is half a period. It is sufficient to show the results of the one-half period and the results of the other half period are symmetric with those of the first half period. The blue area represents $\kappa_r/\kappa_\theta>1$ $(\sigma_r/\sigma_\theta>1)$, which corresponds to the thermal concentration (electric concentration). The yellow area represents $\kappa_r/\kappa_\theta<1$ $(\sigma_r/\sigma_\theta<1)$, which corresponds to the thermal cloak (electric cloak). The red line represents $\kappa_r/\kappa_\theta=1$ $(\sigma_r/\sigma_\theta=1)$, corresponding to the thermal sensor (electric sensor). The shape parameter affects the functional switching of thermal and electric fields. The thermal and electric fields function as concentrators at any time when the shape parameter is $\eta=\ln (r_{i+1}/r_i)/\Delta\theta>1$. The thermal and electric field functions change from concentrator to sensor to cloak over time for shape parameter $\eta<1$. The thermal and electric fields will change from the concentrator to the sensor when $\eta=1$, and the cloak's effect does not appear. Therefore, we can give the shape parameter $\eta$ to theoretically predict what kind of function will appear and when the function will switch. 

\begin{figure}
	\centering
	\includegraphics[width=\linewidth]{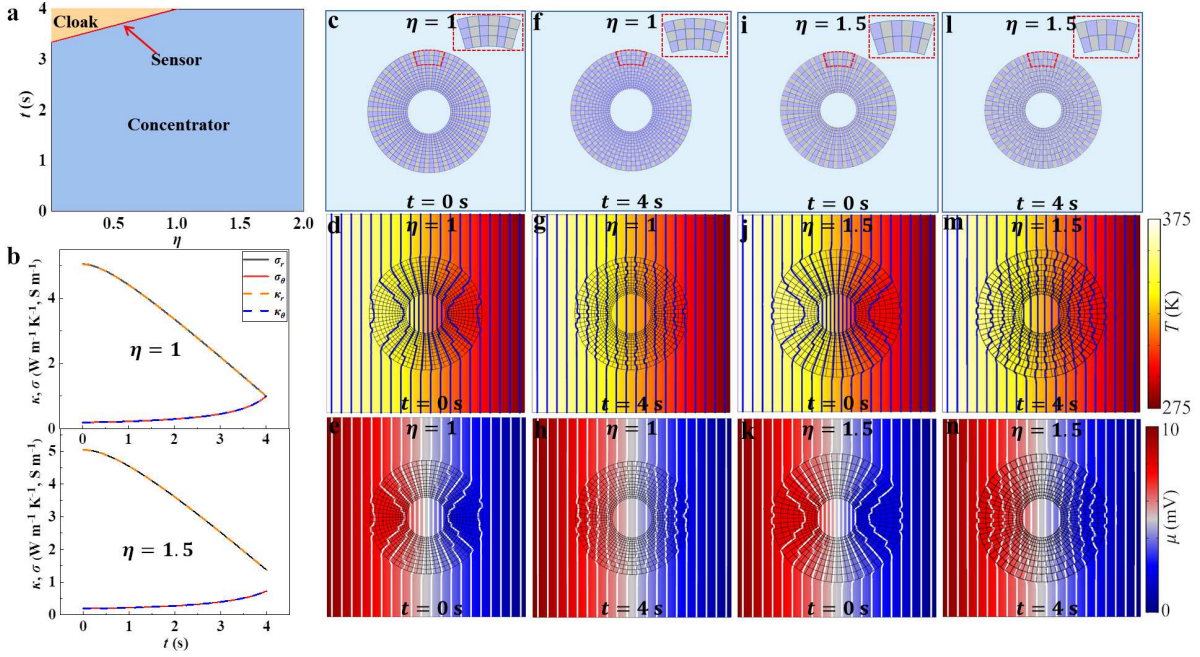}\\
	\caption{Simulation and theoretical results of checkerboard structures with different shapes. (a) Shape and time determine the phase diagrams of thermal and electric functions. (b) Variation of the effective thermal and electric conductivities with $\eta=1$ and $\eta=1.5$. (c)-(h) Structures and simulation results of the checkerboard structure with $\eta=1$ at $t=0~\rm s$ and $t=4~\rm s$. (i)-(n) Checkerboard structure with $\eta=1.5$.}
	\label{fig:boat1}
\end{figure}

Here we select the chessboard structure with the shape parameters $\eta=1$ and $\eta=1.5$, respectively, for simulation. According to the theory, we give the time-dependent changes of the effective thermal and electric conductivity of the two checkerboard structures, as shown in Fig. A.4b. We show the schematic diagram and simulation results of the rotatable checkerboard structure at $t=0~\rm s$ and $t=4~\rm s$ in Fig. A.4c-n. The ratio of the radial thermal (electric) conductivity to the tangential thermal (electric) conductivity of the checkerboard structure with $\eta=1$ is only $\kappa_r/\kappa_\theta=1$ ($\sigma_r/\sigma_\theta=1$) at $t=4~\rm s$, and $\kappa_r/\kappa_\theta=1>1$ ($\sigma_r/\sigma_\theta>1$) at the rest of the time. The simulation results show that the checkerboard structure realizes the function combination of thermal concentration plus electric concentration at $t=0~\rm s$ and thermal sensor plus electric sensor at $t=4~\rm s$. The rotatable checkerboard structure's radial thermal (electric) conductivity with shape parameter $\eta=1.5$ is greater than the tangential thermal (electric) conductivity at any time. Thus, the thermal and electric fields correspond to the concentrator's function at any time. The concentration degree of the concentrators achieved by these two structures is not constant but gradually decreases over time. 


\begin{thebibliography}{99}
	\bibitem{EES21} M.Y. Gao, P. Wang, L.L. Jiang, B.W. Wang, Y. Yao, S. Liu, D.W. Chu, W.L. Cheng, Y.R. Lu, Power generation for wearable systems, Energy Environ. Sci. 14 (2021) 2114-2157, https://doi.org/10.1039/d0ee03911j.
	
	\bibitem{NP22} S.H. Fan, W. Li, Photonics and thermodynamics concepts in radiative cooling, Nat. Photonics 16 (2022) 182-190, https://doi.org/10.1038/s41566-021-00921-9.
	
	\bibitem{MTA22} C.Y. Wang, Z. Vangelatos, C.P. Grigoropoulos, Z. Ma, Micro-engineered architected metamaterials for cell and tissue engineering, Mater. Today Adv. 13 (2022) 100206, https://doi.org/10.1016/j.mtadv.2022.100206.
	
	\bibitem{MTP21-2} Z. Ma, D. Zhao, C. She, Y. Yang, R. Yang, Personal thermal management techniques for thermal comfort and building energy saving, Mater. Today Phys. 20 (2021) 100465, https://doi.org/10.1016/j.mtphys.2021.100465.
	
	\bibitem{AM21} L. Xu, H.Y. Chen, Transformation metamaterials, Adv. Mater. 33 (2021) 2005489, https://doi.org/10.1002/adma.202005489.
	
	\bibitem{MTP22} F. Martinez, M. Maldovan, Metamaterials: Optical, acoustic, elastic, heat, mass, electric, magnetic, and hydrodynamic cloaking, Mater. Today Phys. 27 (2022) 100819, https://doi.org/10.1016/j.mtphys.2022.100819.
	
	
	
	\bibitem{APR22} Y.Z. Shi, Q.H. Song, I. Toftul, T.T. Zhu, Y.F. Yu, W.M. Zhu, D.P. Tsai, Y. Kivshar, A.Q. Liu, Optical manipulation with metamaterial structures, Appl. Phys. Rev. 9 (2022) 031303, https://doi.org/10.1063/5.0091280.
	
	\bibitem{NRM22} H.R. Xue, Y.H. Yang, B.L. Zhang, Topological acoustics, Nat. Rev. Mater. 7 (2022) 974–990, https://doi.org/10.1038/s41578-022-00465-6.
	
	\bibitem{PR21} S. Yang, J. Wang, G.L. Dai, F.B. Yang, J.P. Huang, Controlling macroscopic heat transfer with thermal metamaterials: Theory, experiment and application, Phys. Rep. 908 (2021) 1-65, https://doi.org/10.1016/j.physrep.2020.12.006.
	
	\bibitem{NRP23} Z.R. Zhang, L.J. Xu, T. Qu, M. Lei, Z.K. Lin, X.P. Ouyang, J.H. Jiang, J.P. Huang, Diffusion metamaterials, Nat. Rev. Phys. (2023), in press.
	
	
	\bibitem{PRL19} J. Park, J.R. Youn, Y.S. Song, Hydrodynamic metamaterial cloak for drag-free flow, Phys. Rev. Lett. 123 (2019) 074502, https://doi.org/10.1103/PhysRevLett.123.074502.
	
	\bibitem{PNAS22} M.Y. Chen, X.Y. Shen, Z. Chen, J.H.Y. Lo, Y. Liu, X.L. Xu, Y.L. Wu, L. Xu, Realizing the multifunctional metamaterial for fluid flow in a porous medium, Proc. Natl. Acad. Sci. U.S.A. 119 (2022) e2207630119, https://doi.org/10.1073/pnas.2207630119.
	
	\bibitem{MTP21-1} Y. Xiao, Q.Y. Chen, Q. Hao, Inverse thermal design of nanoporous thin films for thermal cloaking, Mater. Today Phys. 21 (2021) 100477, https://doi.org/10.1016/j.mtphys.2021.100477.
	
	\bibitem{MTP22-1} W. Sha, M. Xiao, M.Z. Huang, L. Gao, Topology-optimized freeform thermal metamaterials for omnidirectionally cloaking sensors, Mater. Today Phys. 28 (2022) 100880, https://doi.org/10.1016/j.mtphys.2022.100880.
	
	\bibitem{MTP22-2} Y.F. Hua, C. Qian, H.S. Chen, H.P. Wang, Experimental topology-optimized cloak for water waves, Mater. Today Phys. 27 (2022) 100754, https://doi.org/10.1016/j.mtphys.2022.100754.
	
	
	\bibitem{ACS19} C. Zhang, W.K. Cao, J. Yang, J.C. Ke, M.Z. Chen, L.T. Wu, Q. Cheng, T.J. Cui, Multiphysical digital coding metamaterials for independent control of broadband electromagnetic and acoustic waves with a large variety of functions, ACS Appl. Mater. Interfaces 11 (2019) 17050−17055, https://doi.org/10.1021/acsami.9b02490.
	
	\bibitem{AOM2020} Y. Zhou, J. Chen, R. Chen, W.J. Chen, Z. Fan, Y.G. Ma, Ultrathin electromagnetic–acoustic amphibious stealth coats, Adv. Opt. Mater. 8 (2020) 2000200, https://doi.org/10.1002/adom.202000200.
	
	
	
	\bibitem{NM19} Y. Li, K.J. Zhu, Y.G. Peng, W. Li, T.Z. Yang, H.X. Xu, H. Chen, X.F. Zhu, S.H. Fan, C.W. Qiu, Thermal meta-device in analogue of zero-index photonics, Nat. Mater. 18 (2019) 48–54, https://doi.org/10.1038/s41563-018-0239-6.
	
	\bibitem{PRAP20} L.J. Xu, G.L. Dai, J.P. Huang, Transformation multithermotics: controlling radiation and conduction simultaneously, Phys. Rev. Appl. 13 (2020) 024063, https://doi.org/10.1103/PhysRevApplied.13.024063.
	
	\bibitem{LiAM20} J.X. Li, Y. Li, P.C. Cao, T.Z. Yang, X.F. Zhu, W.Y. Wang, and C.W. Qiu, A continuously tunable solid-like convective thermal metadevice on the reciprocal line, Adv. Mater. 32 (2020) 2003823, https://doi.org/10.1002/adma.202003823.
	
	\bibitem{IJHMT21} L.J. Xu, J. Wang, G.L. Dai, S. Yang, F.B. Yang, G. Wang, J.P. Huang, Geometric phase, effective conductivity enhancement, and invisibility cloak in thermal convection-conduction, Int. J. Heat Mass Transf. 165 (2021) 120659, https://doi.org/10.1016/j.ijheatmasstransfer.2020.120659.
	
	
	
	\bibitem{JAP2010} J.Y. Li, Y. Gao, J.P. Huang, A bifunctional cloak using transformation media, J. Appl. Phys. 108 (2010) 074504, https://doi.org/10.1063/1.3490226.
	
	\bibitem{PRL2014-3} Y.G. Ma, Y.C. Liu, M. Raza, Y.D. Wang, S.L. He, Experimental demonstration of a multiphysics cloak: manipulating heat flux
	and electric current simultaneously, Phys. Rev. Lett. 113 (2014) 205501, https://doi.org/10.1103/PhysRevLett.113.205501.
	
	\bibitem{PRX2014} M. Moccia, G. Castaldi, S. Savo, Y. Sato, V. Galdi, Independent manipulation of heat and electrical current via bifunctional metamaterials, Phys. Rev. X 4 (2014) 021025, https://doi.org/10.1103/PhysRevX.4.021025.
	
	\bibitem{AM15} T.Z. Yang, X. Bai, D.L. Gao, L.Z. W, B.W. Li, J.T.L. Thong, C.W. Qiu, Invisible sensors: simultaneous sensing and camouflaging in multiphysical fields, Adv. Mater. 27 (2015) 7752–7758, https://doi.org/10.1002/adma.201502513.
	
	\bibitem{OE2015} C.W. Lan, B. Li, J. Zhou, Simultaneously concentrated electric and thermal fields using fan-shaped structure, Opt. Express 23 (2015) 24475-24483, https://doi.org/10.1364/OE.23.024475.
	
	\bibitem{SR2017} T. Stedman, L.M. Woods, Cloaking of thermoelectric transport, Sci. Rep. 7 (2017) 6988, https://doi.org/10.1038/s41598-017-05593-6.
	
	
	
	\bibitem{eLight22} S.X. Yin, E. Galiffi, A. Alù, Floquet metamaterials, eLight 2 (2022) 8, https://doi.org/10.1186/s43593-022-00015-1.
	
	\bibitem{AP22} E. Galiffi, R. Tirole, S.X. Yin, H.N. Li, S. Vezzoli, P.A. Huidobro, M.G. Silveirinha, R. Sapienza, A. Alù, J.B. Pendry, Photonics of time-varying media, Adv. Photonics 4 (2022) 014002, https://doi.org/10.1117/1.AP.4.1.014002.
	
	
	
	\bibitem{AM2019} L. Zhang, X.Q. Chen, R.W. Shao, J.Y. Dai, Q. Cheng, G. Castaldi, V. Galdi, T.J. Cui, Breaking reciprocity with space-time-coding 
	digital metasurfaces, Adv. Mater. 31 (2019) 1904069, https://doi.org/10.1002/adma.201904069.
	
	\bibitem{AP2022-1} S. Taravati, G.V. Eleftheriades, Microwave space-time-modulated metasurfaces, ACS Photonics 9 (2022) 305−318, https://doi.org/10.1021/acsphotonics.1c01041.
	
	
	
	\bibitem{SA21} Z.X. Chen, Y.G. Peng, H.X. Li, J.J. Liu, Y.J. Ding, B. Liang, X.F. Zhu, Y.Q. Lu, J.C. Cheng, A. Alù, Efficient nonreciprocal mode transitions in spatiotemporally modulated acoustic metamaterials, Sci. Adv. 7 (2021) eabj1198, https://doi.org/10.1126/sciadv.abj1198.
	
	\bibitem{AM22} Y.R. Jia, Y.M. Liu, B.L. Hu, W. Xiong, Y.C. Bai, Y. Cheng, D. Wu, X.J. Liu, J. Christensen, Orbital angular momentum multiplexing in space–time 
	thermoacoustic metasurfaces, Adv. Mater. 34 (2022) 2202026, https://doi.org/10.1002/adma.202202026.
	
	
	
	\bibitem{PRL2022-1} L.J. Xu, G.Q. Xu, J.P. Huang, C.W. Qiu, Diffusive fzeau drag in spatiotemporal thermal metamaterials, Phys. Rev. Lett. 128 (2022) 145901, https://doi.org/10.1103/PhysRevLett.128.145901.
	
	\bibitem{PRL2022-2} L.J. Xu, G.Q. Xu, J.X. Li, Y. Li, J.P. Huang, C.W. Qiu, Thermal willis coupling in spatiotemporal diffusive metamaterials, Phys. Rev. Lett. 129 (2022) 155901, https://doi.org/10.1103/PhysRevLett.129.155901.
	
	\bibitem{IJHMT2022} W.X. Zhao, Z. Zhu, Y.W. Fan, W. Xi, R. Hu, X.B. Luo, Temporally-adjustable radiative thermal diode based on
	metal-insulator phase change, Int. J. Heat Mass Transf. 185 (2022) 122443, https://doi.org/10.1016/j.ijheatmasstransfer.2021.122443.
	
	
	\bibitem{AOM2021} H.T. Wu, X.X. Gao, S. Liu, Q. Ma, H.C. Zhang, X. Wan, T.J. Cui, Robust Spin-momentum coupling induced by parity-time 
	symmetric spatiotemporal metasurface, Adv. Opt. Mater. 9 (2021) 210132, https://doi.org/10.1002/adom.202101322.
	
	\bibitem{NC22} W.W. Zhu, H.R.Xue, J.B. Gong, Y.D. Chong, B.L. Zhang, Time-periodic corner states from Floquet higher-order topology, Nat. Commun. 13 (2022) 11, https://doi.org/10.1038/s41467-021-27552-6.
	
	\bibitem{NP2022} G.Q. Xu, Y.H. Yang, X. Zhou, H.S. Chen, A. Alù, C.W. Qiu, Diffusive topological transport in spatiotemporal 
	thermal lattices, Nat. Phys. 18 (2022) 450–456, https://doi.org/10.1038/s41567-021-01493-9.
	
	
	
	\bibitem{AM2017} Z.J. Coppens, J.G. Valentine, Spatial and temporal modulation of thermal emission, Adv. Mater. 29 (2017) 1701275, https://doi.org/10.1002/adma.201701275.
	
	\bibitem{NC2018} L. Zhang, X.Q. Chen, S. Liu, Q. Zhang, J. Zhao, J.Y. Dai, G.D. Bai, X. Wan, Q. Cheng, G. Castaldi, V. Galdi, T.J. Cui, Space-time-coding digital metasurfaces, Nat. Commun. 9 (2018) 4334,  https://doi.org/10.1038/s41467-018-06802-0. 
	
	\bibitem{NC2019} L.L. Li, H.X. Ruan, C. Liu, Y. Li, Y. Shuang, A. Alù, C.W. Qiu, T.J. Cui, Machine-learning reprogrammable metasurface
	imager, Nat. Commun. 10 (2019) 1082, https://doi.org/10.1038/s41467-019-09103-2. 
	
	
	
	\bibitem{P1964} J.B. Keller, A theorem on the conductivity of a composite medium, Physics 5 (1964) 548, https://doi.org/10.1063/1.1704146.
	
	\bibitem{JAP1975} K.S. Mendelson, A theorem on the effective conductivity of a two‐dimensional heterogeneous medium, J. Appl. Phys 46 (1975) 4740, https://doi.org/10.1063/1.321549.
	
	\bibitem{JMP1985} J. Nevard, J.B. Keller, Reciprocal relations for effective conductivities of anisotropic media, J. Math. Phys. 26 (1985) 2761-2765, https://doi.org/10.1063/1.526697.
	
	\bibitem{IJHMT1992} K. Schulgasser, A reciprocal theorem in two-dimensional heat transfer and its implications, Int. J. Heat Mass Transf. 19 (1992) 639, https://doi.org/10.1016/0735-1933(92)90047-L.
	
	
	
	\bibitem{OE2012} S. Guenneau, C. Amra, D. Veynante, Transformation thermodynamics: cloaking and concentrating heat flux, Opt. Express 20 (2012) 8207, https://doi.org/10.1364/OE.20.008207.
	
	\bibitem{PRL2012} S. Narayana, Y. Sato, Heat Flux Manipulation with engineered thermal materials, Phys. Rev. Lett. 108 (2012) 214303, https://doi.org/10.1103/PhysRevLett.108.214303.
	
	\bibitem{SR2013} T. Han, T. Yuan, B. Li, C.W. Qiu, Homogeneous thermal cloak with constant conductivity and tunable heat localization, Sci. Rep. 3 (2013) 1593, https://doi.org/10.1038/srep01593.
	
	
	
	\bibitem{PRAP2019} J.X. Li, Y. Li, T. Li, T.L. Li, W.Y. Wang, L.Q. Li, C.W. Qiu, Doublet thermal metadevice, Phys. Rev. Appl. 11 (2019) 044021, https://doi.org/10.1103/PhysRevApplied.11.044021.	
\end{thebibliography}
\end{document}